\title{On the term ``randomization test''}

\author{Jesse Hemerik\footnote{Biometris, Wageningen University \& Research, P.O. Box 16, 6700 AC Wageningen, The Netherlands. e-mail: jesse.hemerik@wur.nl}}

\documentclass[11pt]{article}

\usepackage{amsmath}
\usepackage{amsfonts}
\usepackage{bm}
\usepackage{bbm}
\usepackage{amsthm}
\usepackage{comment}
\usepackage[nottoc]{tocbibind}
\usepackage{graphicx}

\usepackage{subcaption}

\usepackage{changepage}
\usepackage{natbib}
\usepackage{abstract}
\usepackage[colorlinks=true,citecolor=blue,runcolor=blue,linkcolor=blue, pdfborder={0 0 0}]{hyperref}

\usepackage{tikz}
\usepackage{pgfplots}

\usepackage{algorithm}
\usepackage[noend]{algorithmic}

\usepackage{setspace}
\usepackage[margin=3cm]{geometry}

\begin{document}
\maketitle

\begin{abstract}
\noindent 
There exists no consensus on the meaning of the term ``randomization test''. Contradicting uses of the term are leading to  confusion, misunderstandings and indeed invalid data analyses. As we point out, a main source of the confusion is that the term was not explicitly defined when it was first used in the 1930's. Later authors made clear proposals to reach a consensus regarding the term. This resulted in some level of agreement around the 1970's. However, in the last few decades, the term has often been used in ways that contradict these proposals. This paper provides an overview of the history of the term per se, for the first time tracing it back to 1937. This will hopefully lead to more agreement on terminology and less confusion on the related fundamental concepts.
\\
\\
\emph{keywords:}  history of statistics; Monte Carlo test;  nonparametric inference; permutation test.

\end{abstract}

\section{Introduction}
Nonparametric tests, involving permutations or other rearrangements of data, have a fascinating history, going back to \citet{fisher1935} and earlier if one includes e.g. Fisher's exact test \citep{fisher1934statistical,fisher1935logic} or \citet{eden1933validity}. The most well-known benefit of nonparametric tests is that they do not require distributional assumptions \citep{fisher1935logic,pitman1937significance,welch1937z,good2013permutation}. Another benefit is that in  multi- and high-dimensional settings, simultaneous permutation or transformation may be used to account for complex, unknown dependencies in data \citep{westfall1993resampling,westfall2008multiple,pesarin2010permutation,hemerik2018false,hemerik2019permutation,
blain2022notip,andreella2023permutation}.
Nonparametric tests often require permuting or rearranging a dataset many times, each time computing some statistic based on the transformed dataset. Until the arrival of modern computers, this was often computationally infeasible. This, together with the current importance of high-dimensional data, explains why the last few decades have seen increasing interest in nonparametric inference. 

Nonparametric tests come in various flavours and the term ``randomization tests'' is typically used to refer to some of them.
However,  beyond that, there is no consensus on the meaning of that term.
As e.g. \citet[][p.1272]{arndt1996tests}, \citet[][p.676]{ernst2004permutation}, \citet[][p.121]{berry2014chronicle}, \citet{onghena2018randomization} and \citet[][p.6]{rosenberger2019randomization} note, some authors use the term interchangeably with ``permutation tests''. 
Other authors will use the term for a very broad class that contains permutation-based tests as a special case, while yet others will only consider some permutation-based tests to be randomization tests.
For example, \citet{romano1989bootstrap}, \citet{kennedy1995randomization}, \citet{manly2020randomization} and \citet[][ch.17]{lehmann2022testing} use the term ``randomization test'' in a very general sense, while some  other statisticians  have explicitly argued to use the term in a much more narrow sense \citep{box1953preliminary,kempthorne1969behaviour,gabriel1983rerandomization,ernst2004permutation,edgington2007randomization, onghena2018randomization,rosenberger2019randomization,ramdas2022permutation,zhang2023randomization,liao2023using}. 
These authors propose reserving the term for nonparametric tests based on data from randomized experiments. Such tests have unique properties. 
In particular, if we are not willing to assume that the observations are independent -- e.g., when considering a specific school class -- then such tests still provide valid inference on the sample at hand \citep{pitman1937significance,edgington1966statistical,kempthorne1969behaviour,onghena2018randomization}. 
For example, randomization tests can be validly applied in randomized single-subject experiments \citep{edgington2007randomization}.
 Further,  randomization tests do not always require using a set of transformations that has an algebraic group structure. This leads to some added flexibility in experimental design  \citep{hemerik2021another}.

Due to the inconsistent use of the term ``randomization test'', there has been much confusion regarding differences between types of nonparametric tests. 
This has first of all led to widespread misunderstandings regarding model assumptions and the conclusions that tests allow \citep{edgington1966statistical,kempthorne1969behaviour,edgington1980randomization,ernst2004permutation,onghena2018randomization,rosenberger2019randomization}.
Further, such confusion has led to the use of invalid statistical methods \citep{hemerik2021another}.
Recently statisticians have increasingly felt the need to be explicit about their definition of the term ``randomization test'' and to give some explanation for their choice \citep[in addition to the above authors, see e.g.][]{dobriban2022consistency,ramdas2022permutation,nair2023randomization}.
This is done to acknowledge the fact that there exist different definitions and to reduce confusion among readers.

Some authors, in particular  \citet{onghena2018randomization}, have investigated the history of  the term ``randomization test'' per se, to gain insight into how and why the term was used in the past.
 \citet{onghena2018randomization} traces the term itself back to two publications from 1952 and   notes that the term may likely have been coined by an earlier source.
 However, to the best of our knowledge, no recent author  traces the term farther back than 1952 \citep{david2001annotated,david2008beginnings,berry2014chronicle,onghena2018randomization}.

In this paper,   we trace the term back to multiple publications from the 1930's.
Our aim  is to understand the histories of the existing interpretations of the term ``randomization test''. 
 In particular, we give an overview of the different suggestions that have been made in history  on how to use the term.
 The first usage of the term that we have found is by  E.S. Pearson, who uses the term once in \citet{pearson1937some}. 
 We also discuss the emergence of the related term ``permutation test'', which only happened in the early 1950's. 
 To the best of our knowledge, the early history of that term per se has not been described elsewhere either. For example, \citet{berry2014chronicle} provides an extensive history of permutation tests,  but not of the terminology itself.
 
This paper is built up as follows.
In Section \ref{sechistory}, we discuss the early history of the term ``randomization test''. In particular we discuss the first use of the term in 1937 and the years that follow.
In Section \ref{secptest} we focus on the emergence of the term ``permutation test'' in the 1950's and early discussions on terminology. In particular, we discuss works that explicitly contrast the terms ``randomization test'' and ``permutation test''.
Section \ref{secrecent} discusses the recent history of these terms.

\section{The early history of the term} \label{sechistory}

The written history of the term ``randomization test'' starts with \citet{pearson1937some}, to the best of our knowledge.  Perhaps the term was used earlier in some private communications. E.S. Pearson's paper appeared in Biometrika, of which he had recently become editor, following his father K. Pearson's death \citep{cox2001biometrika}. 
Although \citet{pearson1937some} uses the term only once and does not explicitly provide arguments for using it, his article does provide clues as to why he did.
We first discuss the history leading up to \citet{pearson1937some}, to better understand Pearson's choice of words.

\subsection{Before Pearson's 1937 article}
As is well known, prior to 1937, the term ``randomization'' was commonly used by statisticians in the context of design of experiments \citep{fisher1925statistical, fisher1935}. Further, from 1935 onwards, the knowledge was becoming widespread that this principle of randomization of treatments can be used to construct tests that do not rely on normality \citep{eden1933validity,fisher1935}. For example, \citet{welch1937z} considers nonparametric tests based on what he and several others call ``randomization theory''.

Note   that the term ``permutation test'' was not in use in the 1930's. 
Another important observation to keep in mind is that around 1937, by many statisticians, the prime way to guarantee valid statistical inference was seen to be a randomized experiment, i.e., an experiment involving  randomization of treatments, such as fertilizer. Indeed, \citet[][p.51]{fisher1935} went so far as to say
\\
\\
\emph{
[...] the physical act of randomisation [...] is necessary for the validity of any test of significance [...]}
\\
\\
Such statements probably strengthened readers' association of nonparametric tests with experimental randomization (also see the quote from \citealp{box1953preliminary} below).
Finally, it should be noted that a large part of the data that statisticians faced, came from randomized experiments.
These insights will help to understand why authors in the late 1930's started to use the term ``randomization tests''.

\subsection{Pearson (1937)}
The aim of  \citet{pearson1937some} is to provide an overview of various nonparametric tests based on data from randomized experiments. He calls the nonparametric  test from \citet[][\S21]{fisher1935} a ``test based on randomization''. 
Further in the paper he discusses a different nonparametric test and refers to it as an ``application of this principle of randomization''. These phrases are not surprising, since he  considers tests that rely on experimental randomization.

Keep in mind  that the tests in \citet{pearson1937some}  do not rely on randomly rearranging the dataset, but rearrange the dataset deterministically in all allowed  ways, just like  other early papers that use the term ``randomization test''. 
In other words, these papers do not randomly shuffle data, although this would later become a popular strategy in practice for reducing computation times \citep{hemerik2018exact}. Note that \citet{eden1933validity} do use random shuffling, but do not use the term ``randomization test'' and are not cited by  \citet{pearson1937some}.

One probable reason why \citet{pearson1937some} chose the term is as one might guess:  the test relies on data from experiments involving randomization.
However, to better understand his choice for the term, we must consider the other ways in which he uses the word ``randomization''.
For example, consider his usage of the phrase ``randomization of yields'' \citep[][p.63]{pearson1937some}. There, he is referring to a method in \citet{welch1937z}, which appears in the same number of Biometrika.
\citet{welch1937z} discusses a nonparametric randomization test  for Latin square designs. This test, like  other randomization tests, requires rearranging the  treatment assignments in all allowed ways. This deterministic rearrangement  is what Pearson calls ``randomization of yields'' and several authors would follow this manner of wording.
Part of the explanation for this choice of words may be that the rearrangements that he considers, are analogous to the possible randomization patterns during the physical field experiment. Such an analogy is always present in randomization tests.
For example, consider a standard randomized clinical trial, where 5 individuals receive treatment A and 5 treatment B. Then there are $\binom{10}{5}$ possible treatment randomization patterns, i.e., $\binom{10}{5}$ ways to administer the treatments. Analogously, the randomization test then requires rearranging the data in the $\binom{10}{5}$ corresponding ways.

Another clue to understanding Pearson's choice of words, is his use of the term ``randomization distribution'', which has also been used by several later authors. By this term, he means the distribution obtained by re-computing the test statistic for  all allowed rearrangements of the data. The test compares the observed test statistic with a quantile of this ``randomization distribution''.

To conclude, \citet{pearson1937some} may have chosen the term ``randomization test'' for some of the following four reasons: 1. the fact that he considered data from randomized experiments; 2. the fact that  according to e.g. \citet{fisher1935}, experiments involving randomization were considered the prime way of guaranteeing valid statistical  inference;
3. the analogy between the rearrangements considered by the test and the experimental randomization; 4.  the fact that the test compares an observed statistic with the ``randomization distribution'', in Pearson's words.

\subsection{After 1937}
In the years immediately following 1937, ``randomization test'' is used by a few other authors.
 Both \citet{cochran1938recent} and \citet{mccarthy1939application} call the nonparametric test in \citet{welch1937z} a ``randomization test''. 
\citet{pearson1938some} does not use the term, but he does say ``the distribution [...] under randomization'', which is similar to the term ``randomization distribution'' used in  \citet{pearson1937some}. \citet{nair1940median} writes ``test by randomization''.
Thus, the term  ``randomization test'' was by 1940 already somewhat established,  although it should be added that in the 1940's the term was hardly used.  Instead,  authors preferred to say e.g. ``tests based on the principle of randomization''.

A main question is whether authors saying ``randomization test'' or ``test based on the randomization principle'' associated this terminology with randomized experiments exclusively.
This is difficult to pinpoint, but several early authors did not explictly assume that the data come from a randomized experiment. Here are some examples.
The first one is \citet{neyman1942basic}, which was written in 1939 already. Here, Neyman considers a class of nonparametric tests, that he does not explicitly assume to be based  on data from randomized experiments. 
Instead, he simply assumes ``symmetry of the probability law'' of the variables, which guarantees exactness.
He  nevertheless says that the methodology is ``commonly known as the method of randomization'' and also uses the term ``randomization tests''. 
As further examples, \citet{craig1941note}, \citet{lehmann1949theory}, \citet{wolfowitz1949non}   
 and \citet{hunt1951non} consider  nonparametric tests based on  two samples  from two populations. They do  not explicitly assume that the two samples come from an experiment with randomized treatments.   
They do not use the term ``randomization test'', but, they do all use phrases like  ``test based on the method of randomization''.
The real-life example provided in \citet[][p.29]{lehmann1949theory} is about a randomized experiment, but of course this does not prove that they exclusively had such applications in mind.
\citet{dwass1957} considers tests for two samples from two populations and calls them ``randomization tests'' in the title of his paper. Apart from the title, he does not mention ``randomization'', except when he writes ``randomization tests'' when referring to the methodology in \citet{lehmann1949theory}.
Note that there were certainly also articles on nonparametric testing that did not mention ``randomization'' at all \citep{hotelling1936rank,wald1943exact,wald1944statistical,hoeffding1952large}.

The works by \citet{kempthorne1952design,kempthorne1955randomization}  are the first to \emph{frequently} use the term ``randomization test''. Importantly, unlike the  authors just mentioned,  Kempthorne does explicitly restrict his attention to data from randomized experiments. 
Later,   in \citet[][\S4]{kempthorne1969behaviour} he strongly emphasizes that by ``randomization tests''  he only means tests from such randomized experiments; see \S\ref{secptest}.
Other early users of ``randomization test'' are \citet{anscombe1948validity,kempthorne1953partition,mood1954asymptotic,wilk1955randomization,
siegel1956randomization,napier1956systemic,watson1957sufficient}.
Most of these works are focused on randomized experiments. Some of these works are more mathematical and do not specify whether randomized experiments are considered. \citet{siegel1956randomization} uses the term for a general class of tests that is not restricted to randomized experiments, as is clear from the examples he provides.

\citet{chung1958randomization} do not consider the term to be limited to randomized experiments either. Indeed, they also use the term in the context of observational studies, for example a comparison of alcoholics and non-alcoholics.  
Moreover,  they suggest that some others also use the term in such a general way:
\\
\\
\emph{Tests based on permutations of observations are non-parametric tests. For
the two-sample problem they require no more than the basic assumption of the
problem -- that under the null hypothesis the two samples behave as a single
sample from a population. Actually, they require less  -- that, under the null
hypothesis, the probability distribution is symmetric under all permutations
of the observations. If the populations correspond to different ``treatments,"
this symmetry can be assured by randomly assigning the treatments to the
experimental units. As a result, these tests are often called randomization
tests.}
\\
\\
Thus,  in the interpretation of \citet{chung1958randomization}, experimental randomization is optional and not necessary for a method to be called a randomization test.
On the other hand, they do claim that randomization tests got their name because  they are sometimes based on data from randomized experiments.

\section{Emergence of the term ``permutation tests'' and early discussions on terminology (1950's and 1960's)} \label{secptest} 
To the best our knowledge, the term ``permutation test'' is first used in \citet{cox1952review} and soon after in \citet{kruskal1952use}, \citet{box1953preliminary} and \citet{box1954robust}.
Before 1952, sometimes formulations such as ``tests based on permutations of the observations" were used. For example, \citet{wald1944statistical} and \citet{hoeffding1952large} used this exact wording. It is of course not surprising that at some point this was shortened to ``permutation tests''.
\citet{cox1952review} is a review of the book by \citet{kempthorne1952design}. \citet{cox1952review} uses ``permutation test'' interchangeably with ``randomization test'', although the book under review in fact consistently uses ``randomization test''. Since the randomization tests in Kempthorne were indeed based on permutations, Cox's use of ``permutation test'' was perhaps not considered objectionable.

Around this time, some authors felt that the term ``randomization test'' was strongly associated with experimental randomization. Some thought that this was limiting and that a name was needed for general  permutation-based tests that were not necessarily based on randomized experiments. In this vein, none other than E. S. Pearson'  former PhD student G. E. P. Box, together with S. L. Anderson, wrote in \citet{box1953preliminary}:
\\
\\
\emph{An approach to the problem of hypothesis testing which does not involve assumptions
  concerning the form of the parent distribution was given by R.A. Fisher in 1937} [sic].  \emph{In this method, the argument has   come to be associated with the concept   of randomization (i.e. the concept of arranging an experimental program
  in a randomized design). It should be noted, however, that, in fact, this type of test, known as the randomization   test, might more pertinently be   called a permutation test. In fact it is only necessary to assume that the samples   are drawn from the same distribution. The only restriction on the   distribution is that the likelihood is unchanged   when the observations are rearranged. For example independent observations from any populations,   satisfy this   condition.}
\\
\\
Their point was that  permutation-based testing is not restricted to experimental data, but can also be applied to certain observational data.
This idea was not  new, see e.g. the test in \citet[][p.58]{fisher1936coefficient} comparing Englishmen and Frenchmen. However, \citet{box1953preliminary} felt that permutation-based tests in general had become too much ``associated with the concept  of randomization''. They felt a term was needed for general permutation-based tests that were not necessarily based on randomized experiments. They proposed to use the term ``permutation tests'' for this general class of methods. 

From 1952 onwards, the term ``permutation test'' became increasingly common \citep{box1955permutation,ruist1955comparison,hack1958empirical,wallace1959simplified,lehmann1959testing}, with some authors writing ``permutation or randomization tests'' \citep{box1955permutation,dempster1960significance,cox1963randomization}.
\citet[][\S4]{kempthorne1969behaviour} provides an  insightful discussion on the terminology ``permutation test'' versus ``randomization test''.  They argue \citep[like e.g.][]{edgington1966statistical} that nonparametric tests based on randomized experiments have some unique properties that other permutation-based tests do not; these differences were briefly discussed in the Introduction of this paper. 
For these other permutation-based tests they propose to use the term 
 ``permutation tests''. They propose to reserve the term ``randomization tests'' for tests based on randomized experiments. Similar suggestions are later done in \citet[][p.828]{gabriel1983rerandomization}, \citet[][p.676]{ernst2004permutation}, \citet{edgington2007randomization}, \citet{onghena2018randomization} and \citet[][p.6]{rosenberger2019randomization}. \citet{lehmann1975nonparametrics} always uses ``permutation tests'' and not ``randomization tests'', but does clearly distinguish between the  ``population model'' and the ``randomization model'' \citep[see also e.g.][]{onghena2018randomization}. The former is the well-known model involving independent, uniform sampling from populations. The latter model does not make such assumptions and  requires randomization of treatments. Nonparametric tests based on that model are precisely what e.g. Kempthorne calls randomization tests.

\section{1970's to now} \label{secrecent}
By, say, 1980, the term ``permutation test'' had  become more popular than ``randomization test'' for referring to general tests based on permutations \citep{oden1975arguments,lehmann1975nonparametrics,wellner1979permutation,robinson1982saddlepoint,rosenbaum1984conditional,
tritchler1984inverting}.
 The term ``randomization test'' was still used -- most predominantly for  tests based on data from randomized experiments \citep{zerbe1977randomization,edgington1980randomization,still1981approximate}.

 In the last few decades, the term ``randomization test'' has been increasingly used for very general classes of tests, not necessarily based on randomized experiments
 \citep{bebber2013crop,wang2022approximate}.
 For example, \citet{manly1986randomization,manly1991randomization} uses the term to refer to general tests based on randomly permuting or reordering the data. This interpretation is followed by several others \citep{kennedy1995randomization, jackson1995protest, kennedy1996randomization,depatta1997multivariate,
 peres2001assessing,o2005performance,torres2010properties,
 farine2015constructing,millo2017simple}.

Another example is \citet{romano1989bootstrap, romano1990behavior}. This work considers a broad class of  tests  that use a general group of fixed transformations, with permutations as a special case. 
These general tests are called ``randomization tests'' in \citet{romano1989bootstrap, romano1990behavior}.  
Later works that adopt such a definition of ``randomization test'' are \citet{lehmann2005testing,romano2005exact,canay2017randomization,
dobriban2022consistency}.

Currently,  many authors consider ``randomization tests'' to be a quite general term, that does not necessarily refer to tests involving data from randomized experiments. 
As a recent example, the already classic paper
\citet[][p.557]{candes2018panning} introduces the term ``conditional randomization tests'', which has often been used since. This term refers to a Monte Carlo test that repeatedly randomly samples from a conditional distribution, given certain covariates. 
Such tests do not presuppose  experimental randomization and hence are not randomization tests in the sense of e.g. Edgington and Kempthorne.

\section*{Discussion}
We have seen that the term ``randomization test'' is older than ``permutation test''. The latter term was introduced in the early 1950's for general tests based on permutations. This was done to avoid associations with experimental randomization, since permutation-based tests do not necessarily rely on such randomization. Such distinctions can be important, because tests based on experimental randomization have unique properties, that other permutation methods lack, as discussed in the Introduction.
Unfortunately, there has never been a clear, universally respected consensus regarding the term ``randomization test'' -- although there almost was one around the 1970's.
In the last few decades, the term ``randomization test'' has increasingly been used for  general classes of tests, not necessarily based on experimental randomization. 
The consequence of such contradicting uses of terms is confusion \citep{onghena2018randomization,hemerik2021another}.

\citet{schopenhauer1859welt} said the following on attaching meanings to words:
\\
\\
\emph{The words, therefore, are no longer unappropriated, and to read into them a meaning entirely different from that which they have had hitherto is to misuse
them, to introduce a licence according to which anyone could use
any word in any sense he chose, in which way endless confusion
would inevitably result. Locke has already shown at length that most
disagreements in philosophy arise from a false use of words.}
\\
\\
(translation by \citealp{payne1957translation}). If we agree with this, then  the meaning that early authors like \citet{pearson1937some}, \citet{cochran1938recent} and \citet{neyman1942basic} gave to the term ``randomization test'', is worth studying. What makes this challenging, however, is that these authors mostly  dealt with data from randomized experiments. It is not clear whether some considered certain tests outside this domain to be randomization tests. In any case, what we can do is take note of works from the 1950's and 1960's that explicitly define what is meant and not meant by the term. While \citet{chung1958randomization} do not mind using the term in a very general way,  \citet{box1953preliminary} and \citet[][\S4]{kempthorne1969behaviour} object to this. \citet[][\S4]{kempthorne1969behaviour}  explain  why they consider it important to reserve the term for tests based on experimental randomization. Likewise, \citet{box1953preliminary} propose to call  general permutation-based tests  
``permutation tests''.
Most later authors who  discuss the term ``randomization test'' in detail, roughly agree with the definition of \citet[][\S4]{kempthorne1969behaviour}. Those later authors also emphasize the unique reasoning and properties of tests based on experimental randomization \citep{edgington1980randomization,gabriel1983rerandomization,ernst2004permutation,
onghena2018randomization,rosenberger2019randomization}.

Thus, rather than calling a wide range of approaches ``randomization tests'', using other names can be less confusing. Here are some suggestions. 
Rather than calling all tests based on randomly permuting observational data ``randomization tests'', one could say ``random permutation test''.
As another example, general tests of invariance under a group of transformations \citep{hoeffding1952large,romano1989bootstrap, romano1990behavior} may be called ``group invariance tests'', in line with the existing notion of ``group invariance''
\citep{eaton1989group,giri1996group,romano1990behavior,lehmann2022testing}.

\section*{Acknowledgments}
I thank Patrick Onghena, Cajo ter Braak and Jelle Goeman for valuable comments on an earlier version of this text.

\setlength{\bibsep}{3pt plus 0.3ex}  
\def\bibfont{\small}  

\bibliographystyle{agsm}
\bibliography{bibliography}

\end{document}